\documentclass[aps,twocolumn,showpacs]{revtex4}
\usepackage{amssymb}
\usepackage{natbib}
\usepackage{amsmath}
\usepackage{amsfonts}
\usepackage{graphicx}
\usepackage{mathrsfs}
\usepackage{dcolumn}
\usepackage{bm}
\usepackage{color}
\usepackage{cancel}
\usepackage{mathbbol}
\usepackage{epsfig}
\usepackage{units}

\definecolor{rot}{rgb}{0.75,0.05,0.25}
\definecolor{hellgrau}{gray}{0.5}
\definecolor{blau}{rgb}{0,0,0.7}

\begin{document}
\title{Thermostated Hamiltonian dynamics with log-oscillators}
\author{Michele Campisi}
\email{michele.campisi@physik.uni-augsburg.de}
\affiliation
{Institut f\"ur Physik, Universit\"at Augsburg,
Universit\"atsstrasse 1, D-86135 Augsburg, Germany}
\author{Peter H\"anggi}
\affiliation {Institut f\"ur Physik, Universit\"at Augsburg,
Universit\"atsstrasse 1, D-86135 Augsburg, Germany, and Department of Physics and Centre for Computational
Science and Engineering, National University of Singapore, Singapore 117546}

\begin{abstract}
With this work we present two new methods for the generation of thermostated, manifestly Hamiltonian dynamics and provide corresponding illustrations.
The basis for this new class of thermostats are the peculiar thermodynamics as exhibited by  logarithmic oscillators. These
 two schemes are best suited when applied to  systems with a small number of degrees of freedom.
\end{abstract}

\maketitle
\section{Introduction}
Back in 1984 Nos\'e put forward a method for the generation of equations of motion that sample the canonical ensemble
\cite{Nose84JCP81}.
The method is based on the Nos\'e Hamiltonian, reading:
\begin{equation}
 H= \sum_i \frac{p_i^2}{2m_i X^2} + V(\mathbf x) + \frac{P^2}{2M}+k_BT\ln {X}
\label{Nose}
 \end{equation}
where a log-oscillator with Hamiltonian ${P^2}/{2M}+k_BT\ln {X}
$, is non-linearly coupled to a ``virtual'' system $(\mathbf x, \mathbf p)$.
The thermostated dynamics of the ``real'' system are obtained after a time-rescaling
and the application of a non-canonical transformation. The method was later further developed by Hoover \cite{Hoover85PRA31},
and is currently widely used and known as the Nos\'e-Hoover thermostat.

In this paper we unveil those special thermodynamic properties of log-oscillators which provide them with the  power to act as thermostats and,
based on them, show two more ways in which log-oscillators can be employed to generate thermostated dynamics. At variance with the
method of Nos\'e, these methods are genuinely Hamiltonian, in the sense that the thermostated dynamics are obtained directly from
Hamilton's equations of motion, with no need to perform a time rescaling nor the use of non-canonical transformations \cite{Klages,Kusnezov}.
Consequently these methods not only
constitute a numerical means but, as well,  can even be implemented {\it in situ} with real experiments aimed at thermostating a physical system.
The first of the two methods has been reported  recently with a letter, see in Ref.  \citenum{Campisi12PRL108}. Its feasibility has been further discussed with a short account in Ref. \citenum{Campisi13PRL110}, providing there the response which dispels a criticism raised by Hoover and co-workers \cite{Melendez13PRL110}.

It is important to stress that, just like the Nos\'e-Hoover method, these methods only work provided the overall
dynamics are ergodic, which might present a problem, -- especially when applied to small systems. In the case of Nos\'e-Hoover thermostating
one possible solution was offered by Martyna \emph{et al.} \cite{Martyna92JCP97}, who proposed the use of chains of Nos\'e-
Hoover thermostats.
Our first method,  at least in the implementation we have explored [that is considering a  system of particles which interact with each other and
with a log-oscillator via short range hard core repulsion, see  \ref{eq:H-short} below] seemingly is immune in reference to this ergodicity issue
\cite{Campisi12PRL108,Campisi12UNPUB1,Campisi12UNPUB2}.
Regarding our second method,  see \ref{eq:H-M2} below, the absence of ergodicity may present an issue;  this second  method, however, is sufficiently flexible as to overcome this challenge.
\section{Helmholtz Theorem\label{sec:HT}}
The fact that logarithmic oscillators have a thermostating power
is a consequence of their peculiar thermodynamic properties.
In this section we shall clarify in what sense it is meaningful to talk about
the thermodynamics of mechanical systems that have only one or few degrees
of freedom, -- as it is the case of logarithmic oscillators, and demonstrate
how to calculate their thermodynamic properties.

Our starting point is the salient equation of thermodynamics:
\begin{equation}
\delta Q /T = \text{exact differential} =dS
\end{equation}
also known as the heat theorem \cite{GallavottiBook}.
As early as 1884, Helmholtz proved that this mathematical structure
of thermodynamics is inherent to the classical Hamiltonian dynamics of systems having only one single trajectory
for each energy, which he called monocyclic systems \cite{HelmholtzINBOOK}.
Arguably, this seldom appreciated and rarely known fact was one of the cornerstones on which
ergodic theory (which generalizes Helmholtz monociclicity) and statistical mechanics were later built up by Boltzmann and others
\cite{GallavottiBook,Campisi05SHPMP36,Campisi10AJP78,Hertz10AP338a}.

The Helmholtz theorem goes as follows: Consider a classical particle in a confining
potential $\varphi(X;\lambda)$, where $\lambda$ is an external parameter.
To each couple ($E,\lambda$) of values of the energy and the external parameter is associated one
closed trajectory in the system phase space. For each trajectory one can calculate the
average quantities:
\begin{align}
k_B T(E,\lambda) &:= \left \langle \frac{P^2}{M}  \right \rangle_{E,\lambda} \label{eq:TElambda}\\
F(E,\lambda) &:= - \left \langle \frac{\partial \varphi}{\partial \lambda} \right \rangle_{E,\lambda}\label{eq:PElambda}
\end{align}
where $P,M$ are the particle momentum and mass respectively, and $\langle \cdot \rangle_{E,\lambda}$
denotes time average over the trajectory specified by $(E,\lambda)$.
Noticing that $F(E,\lambda)$ is the average force that the particle exerts against the external agent, keeping the parameter
$\lambda$ at a fixed value, one realizes that
\begin{align}
\delta Q = dE + F(E,\lambda)d\lambda
\end{align}
represents the heat differential. The Helmholtz theorem
states that $1/T(E,\lambda)$ is an integrating factor for $\delta Q$,
\begin{equation}
\frac{dE + F(E,\lambda)d\lambda}{T(E,\lambda)} = \text{exact differential} =dS
\label{eq:heat-theo}
\end{equation}
and that
\begin{equation}
S(E,\lambda) = k_B \ln \Phi(E,\lambda)
\label{eq:S}
\end{equation}
where
\begin{align}
\Phi(E,\lambda)&=\left[2 \int_{X_-(E,\lambda)}^{X_+(E,\lambda)} \sqrt{2M(E-\varphi(X;\lambda))}dX/h \right]\nonumber	\\
&= \int dX dP\,  \theta[E-H(X,P)]\, .
\end{align}
Here, ${X_\pm(E,\lambda)}$ are the turning points of the trajectory, $h$ is a constant with the units of an action,
and $\theta(x)$ denotes Heaviside step function.
Accordingly it is meaningful to call $T(E,\lambda)$ the temperature of the particle and
$S(E,\lambda)$ its entropy. $S(E,\lambda)$ in \ref{eq:S} is also known as the Hertz entropy \cite{Hertz10AP338a}.

Once the function $S(E,\lambda)$ is known, one can then quickly calculate $T(E,\lambda)$ and $F(E,\lambda)$ in accordance to \ref{eq:heat-theo}, as:
\begin{align}
T &= \left( \frac{\partial S}{\partial E} \right)^{-1}\\
F &=\frac{\partial S}{\partial \lambda}
\left( \frac{\partial S}{\partial E} \right)^{-1}
\end{align}
and so obtain the thermodynamics of the system: equation of state, specific heat, etc..

Following this scheme, in the next section, we will proceed to derive the thermodynamics of log-oscillators and
highlight the peculiar properties that provide them with thermostating power.

\section{The peculiar thermodynamics of a log-oscillator}
\label{sec:TD}

\subsection{The heat capacity is infinite}
\begin{figure}[]
		\includegraphics[width=.5\textwidth]{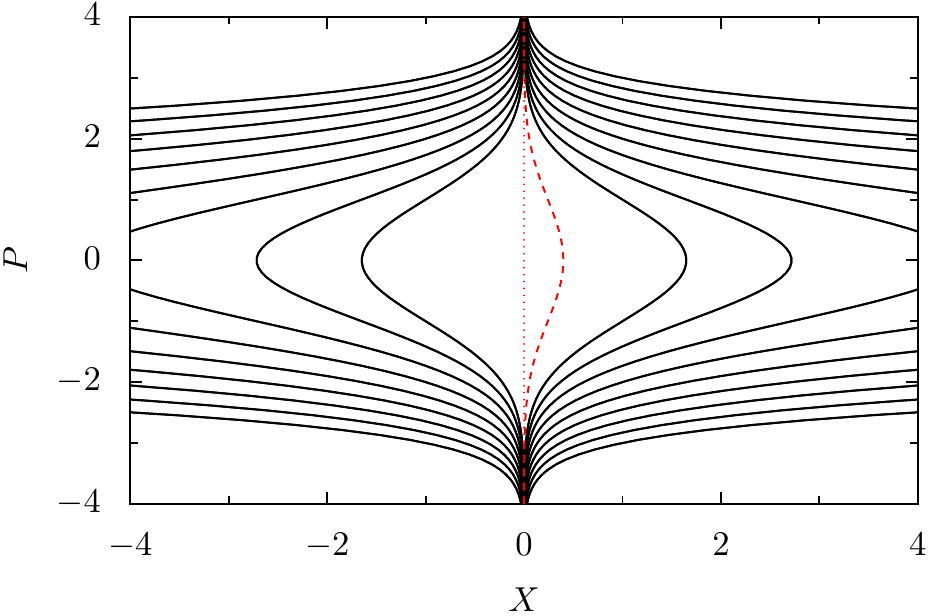}
		\caption{(Color online) Black solid lines: Phase space trajectories of a log-oscillator at energies $E=1/2, 1, 3/2,  \dots 9/2$, inner curves have lower energies. Red dashed line: the momentum distribution function, \ref{eq:f(P)}. Here $M=1$, $k_BT=1$.
		}
		\label{fig:Fig1}
\end{figure}

Let us consider a log-oscillator with Hamiltonian:
\begin{align}
H_{\text{log}}(X,P) = \frac{P^2}{2M}+k_BT\ln\frac{|X|}{b}\, ,
\label{eq:Hlog}
\end{align}
where $M$ is the mass and $b$ some positive constant with the dimension of length.
Figure \ref{fig:Fig1} depicts some trajectories in phase space of different energies.
Solving the equation $H_{\text{log}}(X,P)=E$ for $X$, one sees that the trajectories are given
by the equations:
\begin{equation}
X= \pm \, b\,  e^{E/k_BT} e^{-P^2/2Mk_BT}\;.
\end{equation}
That is the trajectories possess a Gaussian shape. Note that, accordingly, the maximal excursion grows exponentially with $E/k_BT$:
$X_{max}= b\,  e^{E/k_BT}$.
A straightforward calculation gives:
\begin{align}
\Phi_\text{log}(E)= \int dX dP\,  \theta[E-H_{\text{log}}(X,P)] = 2b \sqrt{2\pi M k_B T} e^{E/k_BT}\, .
\end{align}
Here, and in what follows we have set for convenience $h=1$.
Accordingly, the entropy, \ref{eq:S}, reads:
\begin{align}
S(E)&= \frac{E}{T} + k_B  \ln  [2b \sqrt{2\pi M k_B T}].
\end{align}
Using the Helmholtz theorem, we get:
\begin{align}
\langle P^2/M\rangle_E= (\partial S/ \partial E)^{-1} = k_BT.
\label{eq:T(E)=T}
\end{align}
This expresses the major feature of the thermodynamics of a log-oscillator:
all its trajectories inherit one and the same absolute temperature, which is given by
$T$, where $T$ is the strength of the logarithmic potential.
This fact is very peculiar: Consider for example the 1D harmonic oscillator,
in this case $k_B T(E) = E$, namely the higher the energy, the higher the temperature.
Similarly this is the case for a particle in a 1D box, where  $k_B T(E)= E/2$.

It therefore follows that the log-oscillator possesses a spectacular property:  
it has an {\it infinite} heat capacity; i.e.,
\begin{align}
C(E)&= (\partial T/ \partial E)^{-1} = \infty
\end{align}
thus, it mimics  a bath composed of an infinite
collection of harmonic oscillators \cite{Chaos2005}, or one with an infinite number of particles in a box.

\subsection{Log-oscillators sample the Maxwell distribution}
Yet another peculiar feature of the log-oscillator is that
the probability density $f(P)$ to find it with momentum $P$,
is given by the Maxwell distribution at temperature $T$:
\begin{align}
f(P)= (2\pi M k_BT)^{-1/2}e^{-P^2/2Mk_BT}\;.
\label{eq:f(P)}
\end{align}
This holds independent of its energy $E$.
To see this, consider the trajectory of the log-oscillator of some energy $E$.
The probability to find the system at $X,P$, is given by the
microcanonical distribution:
\begin{align}
\rho(X,P)= \delta[E-H_\text{log}(X,P)]/\Omega_\text{log}(E)
\end{align}
where $\delta(x)$ denotes Dirac's delta function, and
\begin{align}
 \Omega_{\text{log}}(E) &= \int dX dP\,  \delta[E-H_{\text{log}}(X,P)] \nonumber \\
 &= \frac{\partial \Phi_\text{log}(E)}{\partial E} = 2 b\sqrt{2 \pi M/k_BT}\,  e^{E/k_BT}\, .
\label{eq:Omega-log}
\end{align}
Therefore, the probability to find the log-oscillator at momentum $P$,
is obtained by the marginal distribution:
\begin{align}
f(P)&= \int dX \rho(X,P)= \int dX \delta[E-H_\text{log}(X,P)]/\Omega_\text{log}(E) \nonumber \\
&=\frac{1}{\Omega_\text{log}(E)} \frac{\partial}{\partial E}\int dX \theta[E-P^2/2M -k_BT\ln \frac{|X|}{b}] \nonumber \\
&= \frac{2}{\Omega_\text{log}(E)} \frac{\partial}{\partial E}\int_{0}^{b \exp[(E-P^2/2M)k_BT]} dX \nonumber \\
&= \frac{2 }{\Omega_\text{log}(E)} \frac{\partial}{\partial E} b e^{(E-P^2/2M)/k_BT}
= \frac{e^{-P^2/2Mk_BT}}{\sqrt{2\pi Mk_B T}}
\end{align}
where we have used $\delta(y)= d\theta(y)/dy$.

From \ref{eq:f(P)} it is immediate to obtain
that $T(E)=\langle P^2 \rangle_E/Mk_B =T$, in accordance with \ref{eq:T(E)=T}.

The red dashed curve  in Fig. \ref{fig:Fig1} illustrates \ref{eq:f(P)}. When projecting the microcanonical
distribution of the log-oscillator onto the $P$ axis, the Maxwell distribution
is obtained, regardless of the energy.

\section{Method I}
\label{sec:M1}
The central feature of a thermal bath is that its heat capacity is infinite,
hence, in this sense a single log-oscillator does indeed act like a thermal bath.
Based on this observation it is reasonable to expect that
when a system interacts weakly with a log-oscillator, the latter
should induce thermostated dynamics at temperature $T$ in the system.

That this is indeed the case can be seen formally in the following manner \cite{Campisi12PRL108}.
Consider the total Hamiltonian:
\begin{align}
H(\mathbf x,\mathbf p,X,P)= H_S(\mathbf x,\mathbf p)+ H_\text{log}(X,P) + h(\mathbf x, X)
\label{eq:H-short}
\end{align}
where
\begin{align}
H_S(\mathbf x,\mathbf p)= \mathbf p^2/2m + U(\mathbf x)
\label{eq:HS}
\end{align}
is the system Hamiltonian, and $h(\mathbf x, X)$
is a {\it weak} interaction term that couples the system to the log-oscillator.
Under the assumption that the total dynamics are \emph{ergodic}, the
probability density function $p(\mathbf x,\mathbf p)$ for finding the system at $(\mathbf x,\mathbf p)$
reads \cite{Khinchin49Book}:
\begin{align}
p(\mathbf x,\mathbf p) = \frac{\Omega_\text{log}[E_\text{tot}-H_S(\mathbf x,\mathbf p)]}{\Omega(E_\text{tot})}
\label{eq:pSys}
\end{align}
where $E_{\text{tot}}$ is the total energy of the compound system and
\begin{equation}
 \Omega(E_{\text{tot}}) = \int \mathrm{d}{X}
\mathrm{d}{P}\mathrm{d}\mathbf{x}\mathrm{d}\mathbf{p}\,
\delta[E_{\text{tot}}-H(\mathbf{x},\mathbf{p},X,P)]
\end{equation}
is the density of states of the compound system.
Note that the shape of the distribution $p(\mathbf x,\mathbf p)$ is given by the
numerator, whereas the denominator only represents a normalization factor.
Thus, from the fact that the density of states of a log-oscillator is exponential in  ${E/k_BT}$, see
\ref{eq:Omega-log}, it immediately follows that:
\begin{align}
p(\mathbf x,\mathbf p) = \frac{e^{-H_S(\mathbf x,\mathbf p)/k_BT}}{Z(T)}
\label{eq:gibbs}
\end{align}
where $Z(T)=\int d\mathbf x d\mathbf p\,  e^{-H_S(\mathbf x,\mathbf p)/k_BT}$.
Thus, the constant temperature equations of motion read:
\begin{displaymath}
\left\{ \begin{array}{lll}
\dot{\mathbf x} = \mathbf {p}/m, \\
\dot{\mathbf p} = -\partial_{\mathbf x} U(\mathbf x) -\partial_{\mathbf x} h(\mathbf x,X) \\
\dot X = P/M\\
\dot P = -k_BT/X -\partial_X  h(\mathbf x,X)
\end{array}\right.
\end{displaymath}
where $\partial_{\mathbf x}$ denotes the gradient operator in the $\mathbf x$ space
and $\partial_X$ is a short notation for $\partial/\partial X$.
Note that for $h=0$, i.e., in absence of interaction, the system undergoes
constant energy dynamics.

\subsection*{Illustration}
Ref. \citenum{Campisi12PRL108} illustrates the numerical implementation of this method for small systems composed of few particles contained
in a box and interacting through a repulsive hard core potential
\begin{equation}
V_{LJ}(q)=
 \left\{
  \begin{array}{ll}
0\, , &  |q| > 2^{1/6}\sigma \\
4\varepsilon \left[\left(\frac{\sigma}{q}\right)^{12}-\left(\frac{\sigma}{q}
\right)^ { 6 } \right ] +\varepsilon\, , & |q| < 2^{1/6}\sigma
 \end{array}
 \right. \, ,
\label{eq:V(x)}
\end{equation}
The main limitation of this method comes from the fact that, in practical realizations, the logarithmic potential needs to be
truncated at low values of $X$, for example by substituting it with:
 \begin{equation}
 \varphi_{b}(X)= \frac{k_BT}{2}\ln\frac{X^2+b^2}{b^2} \, .
 \label{eq:phi(x)}
\end{equation}
This truncation results in a deviation of the single particle velocity distribution from the target Maxwell distribution. This deviation becomes more and
more pronounced as the number of particles in the system increases, see Fig. 3 of Ref. \citenum{Campisi12PRL108},
and can be compensated by rising the system energy as $E \sim fk_BT/2$, where $f$ is the number of degrees of freedom of the system.
This energy rising, however, is accompanied by an exponential increase of the corresponding length and time scales involved in the
dynamics which go as $e^{E/k_BT} \sim e^{f/2}$, thus limiting the applicability of the method to systems with a \emph{small} number of degrees of freedom.

A prominent novel aspect of this method when compared to the other existing methods discussed in the literature is that it can be implemented not only with
computer simulations but also in analogue simulations, provided one is able to implement the Hamiltonian in \ref{eq:H-short} in a real
experiment \cite{Campisi12PRL108}. Reference \citenum{Campisi13PRL110} discusses such an experimental feasibility of this method using cold
atoms and laser fields.

\begin{figure}[t]
		\includegraphics[width=.5\textwidth]{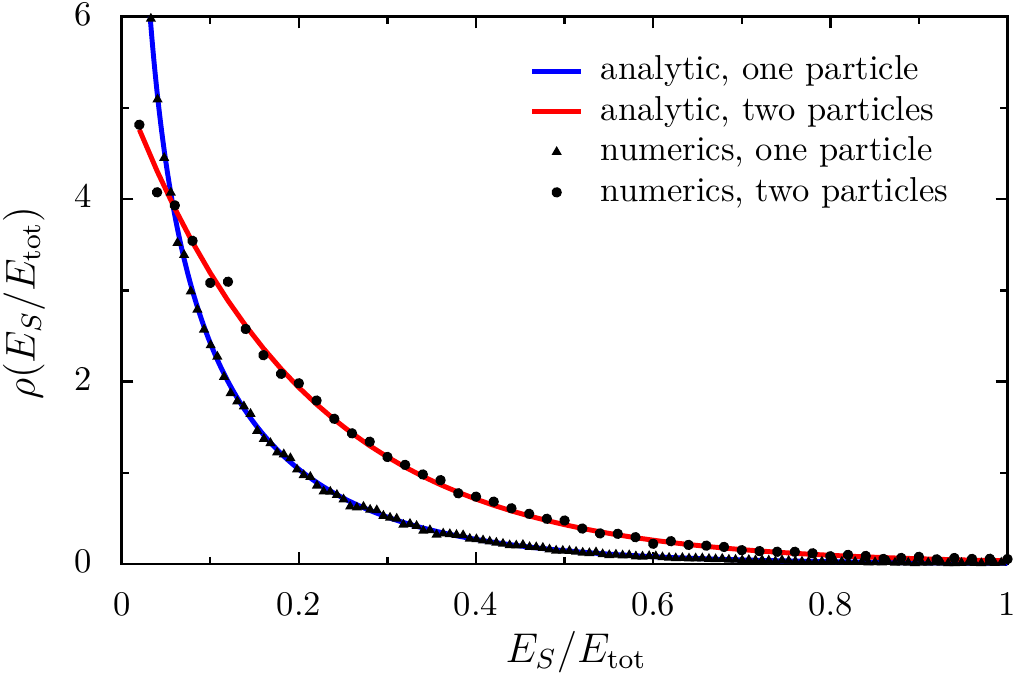}
		\caption{(Color online) Illustration of Method I. Normalized probability density function of energy for a system of $n$ particles in a 1D box performing short ranged collisions, \ref{eq:V(x)}, with a truncated log-oscillator, \ref{eq:phi(x)}, of strength $k_BT = 15 \varepsilon$. The total simulation energy is $E_\text{tot}=5 k_BT$, the box length is $L=10e^{E_\text{tot}/k_BT} \sigma \simeq 1484 \sigma$ and the log-oscillator cutoff length was set to $b=\sigma$. Black triangles: numerical simulation with $n=1$. Black dots: numerical simulation with $n=2$. Blue line: Gibbs distribution at temperature $k_BT=15 \varepsilon$ for $n=1$, as it follows from \ref{eq:gibbs2}.
Red line: Corresponding Gibbs distribution at temperature $k_BT=15 \varepsilon$ for $n=2$ as it follows from \ref{eq:gibbs2}.  This Figure has been provided by Fei Zhan and is adapted here from our Ref. \citenum{Campisi12UNPUB2}.}
		\label{fig:method1}
\end{figure}

Figure \ref{fig:method1} illustrates this method for a system composed of either one particle or two particles in a one-dimensional box performing short range, hard core collisions, \ref{eq:V(x)}, with the truncated log-oscillator in \ref{eq:phi(x)}. It reports the probability $\rho(E_S)$ to find the particle at energy $E_S$ during a long simulation run.
 A symplectic integrator was used to produce the trajectory of the total system and the initial condition was sampled randomly from the shell $E_\text{tot}= 5 k_BT$. 
The numerically computed probability (relative frequency) $\rho(E_S)$ is compared to the expected Gibbs distribution calculated from \ref{eq:gibbs} according to the standard rules of probability theory as
\begin{equation}
 \rho(E_S)=\frac{e^{-E_S/T}\Omega_S(E_S) }{Z(T)} = \frac{e^{-E_S/T}\Omega_S(E_S) }{\int_0^\infty e^{-E_S/T}\Omega_S(E_S) dE_S} \, ,
\label{eq:gibbs2}
\end{equation}
where $\Omega_S(E_S)$ is the density of states of the system. In calculating it we neglect the contribution coming from the short range interaction, thus obtaining  $\Omega_S(E_S) \propto E_S^{n/2-1}$, with $n=1,2$ being the number of particles in the system. For $n=1$ this yields $\Omega_S(E_S)\propto E_S^{-1/2}$ while for $n=2$ we find that $\Omega_S(E_S)$ is a constant. The agreement between theory and simulations  is excellent. Further details and discussion can be found in Refs. \citenum{Campisi12PRL108} and \citenum{Campisi12UNPUB2}.

\section{Method II}
\label{sec:M2}
An alternative method to produce thermostated dynamics is to
couple the system to a free particle via a logarithmic interaction potential.
More explicitly, the statement is that the extended Hamiltonian
\begin{align}
H(\mathbf x,\mathbf p,X,P) &= H_S(\mathbf x,\mathbf p)+ P^2/2M+ k_BT \ln\left(|g(\mathbf x, \mathbf p)-X|/b\right)
\label{eq:H-M2}
\end{align}
produces thermostated system dynamics, provided the (otherwise arbitrary)
function $g(\mathbf x, \mathbf p)$ induces \emph{ergodic} dynamics of the total system.

To demonstrate this, consider the probability $\rho(\mathbf x, \mathbf p, X, P)$ to find the total system at
$(\mathbf x, \mathbf p, X, P)$. Thanks to the ergodic assumption, this is given by
the microcanonical distribution
\begin{align}
\rho(\mathbf x, \mathbf p, X, P) = \delta[E_{\text{tot}}-H(\mathbf{x},\mathbf{p},X,P)]/\Omega(E_{\text{tot}})\, ,
\end{align}
hence:
\begin{align}
p(\mathbf x, \mathbf p) = \frac{\int dX dP\delta[E_{\text{tot}}-H_S-P^2/2M- T \ln(|g-X|/b)]}{ \Omega(E_{\text{tot}})}
\end{align}
Making the change of variable $X'= X-g(\mathbf x, \mathbf p)$, one obtains,
{\it irrespective} of $g(\mathbf x, \mathbf p)$
\begin{align}
p(\mathbf x, \mathbf p) = \frac{\int dX' dP\delta[E_{\text{tot}}-H_S-P^2/2M- k_BT \ln(|X'|/b)]}{\Omega(E_{\text{tot}})}
\end{align}
Note that the numerator is the log-oscillator density of states $\Omega_\text{log}$ taken at $E_{\text{tot}}-H_S$.
Therefore, just as with Method I:
\begin{align}
p(\mathbf x, \mathbf p) =  \frac{\Omega_\text{log}[E_\text{tot}-H_S(\mathbf x,\mathbf p)]}{\Omega(E_\text{tot})}= \frac{e^{-H_S(\mathbf x,
\mathbf p)/k_BT}}{Z(T)}.
\end{align}

The constant temperature equations of motion of this second method read:
\begin{displaymath}
\left\{ \begin{array}{lll}
\dot{\mathbf x} = \mathbf {p}/m  + k_BT[g(\mathbf x, \mathbf p)-X]^{-1}\partial_{\mathbf p} g(\mathbf x, \mathbf p)\\
\dot{\mathbf p} = -\partial_{\mathbf x} U(\mathbf x, \mathbf p) - k_BT[g(\mathbf x, \mathbf p)-X]^{-1}\partial_{\mathbf x} g(\mathbf x, \mathbf
p) \\
\dot X = P/M\\
\dot P = k_BT[g(\mathbf x, \mathbf p)-X]^{-1}
\end{array}\right.
\end{displaymath}
note that for $T=0$ the system undergoes constant energy dynamics.

It is important to repeat that thermostated system dynamics are only reached if the
global dynamics are ergodic. As illustrated below, this requirement is however not too restrictive,
because we have the freedom to chose the function $g(\mathbf x, \mathbf p)$.

\subsection*{Illustration}
To illustrate the method we considered a quartic oscillator:
\begin{equation}
H_S= p^2/2m + k x^4/4
\end{equation}
We simulated the compound system
dynamics using a symplectic integrator with a time step $\Delta t= 10^{-2}b \sqrt{M/k_BT}$ for
a total simulation time $\mathcal T =  1.287 \times 10^9 \Delta t $ time steps.
In our simulations we set $k_B T, b$ and $ M$ as units of energy, length and mass, respectively. We took
$(x_0,X_0,p_0,P_0)=(2,-1,1,-1) $ as the initial condition, $k= k_BT b^{-4}$, and $m=M$.
We computed the probability distribution function $\rho(E_S)$ to find the system at energy
$E_S$, and compared it with the target Gibbs distribution, \ref{eq:gibbs2}. The latter reads
\begin{equation}
\rho(E_S) =  \frac{e^{-E_S/k_BT} E_S^{-1/4}}{\int_0^\infty \mathrm{d}E_S e^{- E_S/k_BT} E_S^{-1/4}}
\end{equation}
where the factor $E_S^{-1/4}$ stems from the density of states of the quartic oscillator:
$\int \mathrm{d}x\mathrm{d}p \delta[E_S-p^2/2m + k x^4/4] \propto E_S^{-1/4}$.
We further computed the probability distribution function to find the system with a velocity of modulus
$v$, and compared it to the target Maxwell distribution, reading:
\begin{equation}
p(v) =  \frac{e^{-mv^2/2k_BT}}{\int_0^\infty e^{-mv^2/2k_BT}}
\end{equation}

Following Ref. \citenum{Campisi12PRL108}, the numerical evaluation of
$\rho(E_S)$ proceeded by recording the value of $E_S$, once every 100 time steps. We divided the energy interval
$[0,E_{tot}]$ in 50 bins, and counted how many times $E_S$ was within each bin, so as to construct a histogram, which,
after normalization gives an approximation to the actual $\rho(E_S)$. A similar procedure was followed for the
calculation of $p(v)$.

\begin{figure}
		\includegraphics[width=.4\textwidth]{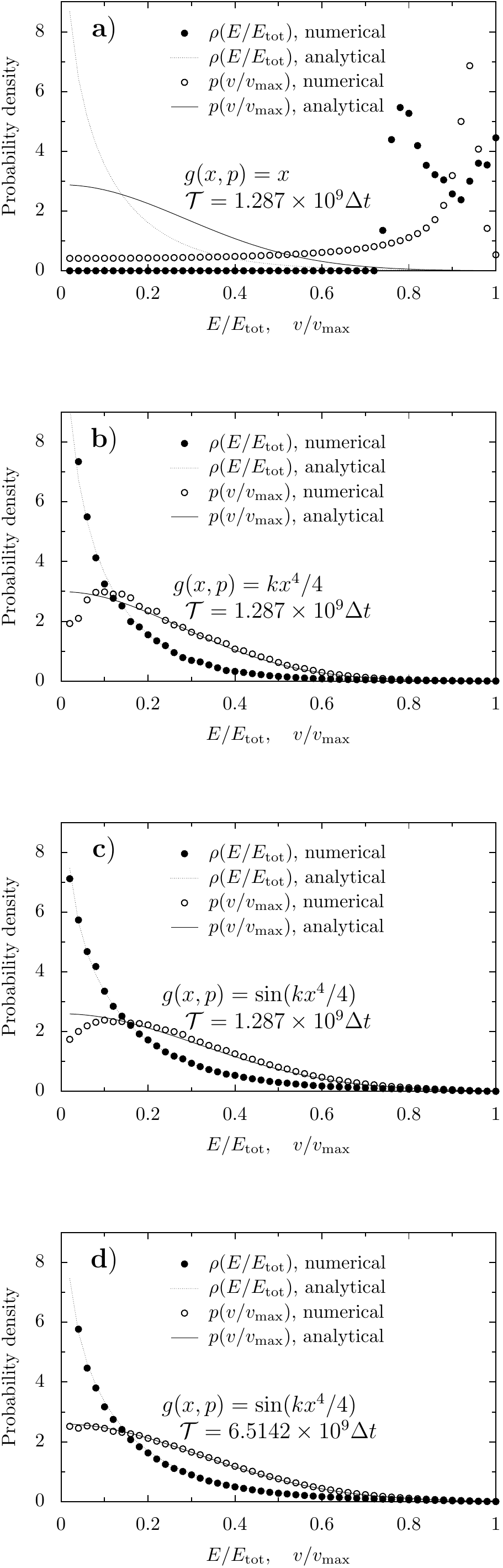}
		\caption{Illustration of Method II. Each panel reports the analytically and numerically computed
		probability distribution for system energy $E_S$ (rescaled by the total fixed energy $E_{\rm{tot}}$)
		and system speed $v$ (rescaled by the maximal speed $v_{\rm{max}}=\sqrt{2E_{\rm{tot}}/m}$),
		for various choices of $g(x,p)$. Panel d) has the same $g(x,p)$ as panel c), but for a longer simulation time.
		}
		\label{fig:method2}
\end{figure}

To begin with we chose $g(x,p)=x$. Notwithstanding the long integration time, the method
fails to converge to the desired target distributions, see Fig. \ref{fig:method2}, panel a).
This means that  with the choice of $g(x,p)=x$, the overall dynamics is not sufficiently ergodic to make the system sample
the canonical ensemble.

The ergodicity of the overall dynamics can be improved by choosing a different form for the
function $g(x,p)$.
Panel b) of  Fig. \ref{fig:method2} reports the result of a dynamical simulation of the same system as in panel a), with the same time-step
$\Delta t$ and simulation time, but with  $g(x,p)= k x^4/4$, namely we chose $g(x,p)$ as the system
potential energy.
While we found a very good agreement between the computed energy probability distribution function and the Gibbs
distribution, the
agreement between the computed absolute velocity distribution and the target Maxwell distribution is
still not very good.
With $g(x,p)=\sin(k x^4/4)$, see in panel c) of Fig. \ref{fig:method2}, reasonably good agreement between simulation and
Maxwell distributions was achieved, while the agreement between the energy distribution and
the Gibbs distribution is excellent. Excellent agreement is achieved with longer simulation times, see in panel d) of Fig. 3.

\section{Remarks}
As emphasized above, ergodicity  of the global dynamics constitutes the crucial  prerequisite for the presented methods to work properly.
Ergodicity suffices and no stronger condition, e.g., the system being mixing, \cite{Lebowitz73PT26} is necessary because all that is needed for the system to sample the Gibbs distribution is that the compound system samples the microcanonical distribution.
It should also be mentioned that ergodicity is a sufficient but not necessary condition for the methods to work, namely in some cases the methods might work even if ergodicity does not hold.

In Method I, whether ergodicity holds depends on the specific choice of the interaction energy $h(\mathbf x, X)$,
which must be chosen in any case weak. In Ref. \citenum{Campisi12PRL108} $h(\mathbf x, X)$ was chosen as a hard-core, short range repulsive interparticle potential, \ref{eq:V(x)}, and that was sufficient for achieving thermostating.
In Method II, the ergodicity property depends on the choice of $g(\mathbf x, \mathbf p)$, which in turn fixes the interaction term $k_BT\ln |
g(\mathbf x, \mathbf p)-X|$. It must be emphasized however that our analysis does neither show formally nor numerically that the total dynamics are indeed ergodic in the examples presented, but only that, loosely speaking, the system appears ``ergodic enough''
for the methods to work.

Note that in method II  the interaction term $k_BT\ln | g(\mathbf x, \mathbf p)-X|$ gives rise to long-range forces. So at variance with the implementation of Method I in Ref.
\citenum{Campisi12PRL108}, where the system and the ``bath'' interacted sporadically through almost instantaneous collisions,
in Method II they constantly influence each other, due to the long range force. 

We have shown how different choices of $g(\mathbf x, \mathbf p)$ can result in different ergodic properties of Method II. An
important subject for further studies would be to derive a set of criteria for appropriately choosing $g(\mathbf x, \mathbf p)$, given the properties of
the system, as encoded in its Hamiltonian $H_S(\mathbf x, \mathbf p)$.

Besides choosing $h(\mathbf x, X)$ or $g(\mathbf x, \mathbf p)$, the ergodicity of both methods can be improved also by substituting the
log-oscillator with a multi-dimensional log-oscillator, which will add more degrees of freedom to the whole system, see in Appendix.

In implementing Method II, we have replaced the logarithmic potential with the same truncated potential, \ref{eq:phi(x)}, used for Method I.
Therefore, just as with Method I, this truncation can lead to deviations to the target Maxwell distribution when the number of particles in the
system increases. An interesting line for future studies would then be to put forward implementations that avoid the truncation and treat the singularity by some other means, which might allow for applying the methods to large systems as well.

\section{Conclusions}
With this study we presented two  Hamiltonian schemes which allow a system $H_S$ to sample a canonical Gibbs distribution. This being so, the method of thermostating is achieved here in a deterministic time-reversal invariant and symplectic manner. Both  schemes rest upon the spectacular thermodynamic property of logarithmic oscillators of  having an infinite heat capacity. Hence, in our methods a single log-oscillator substitutes an infinite heat bath coupled weakly to the system. With our Method I we couple the system weakly to a log-oscillator where the absolute temperature $T$ denotes the strength of the logarithmic potential. In Method II we consider a composite system of $H_S$ and a {\it free} particle which is coupled with a long range log-interaction of strength $T$ to the system of interest $H_S$. Note that Gibbs thermalization occurs here independent of the interaction-strength $T$, being either strong (large $T$) or weak (small $T$). A prominent property inherent to both schemes is that these are  manifestly Hamiltonian \cite{Campisi12UNPUB1}. Also, at variance with the Nos\'e Hamiltonian, \ref{Nose},  our Hamilton functions possess standard (i.e. coordinate-independent) kinetic energy contributions. This fact in turn allows not only an implementation with numerical means but as well a physical realization. This advantage should be contrasted nevertheless with the limitation that both methods inherit from performing a truncation of the logarithmic potential as in \ref{eq:phi(x)}, which, as thoroughly emphasized in our previous accounts \cite{Campisi12PRL108,Campisi13PRL110}, limits an efficient thermostating to systems with a {\it small} number of degrees of freedom. Notably, the investigation of such nano-scale systems is in the limelight of present day research activities \cite{Campisi11RMP83,Jarzynski11ARCMP2,Seifert08EPJB64,Bloch08RMP80}.

\section*{Acknowledgement}This work was supported by by  the German Excellence Initiative ``Nanosystems Initiative
Munich (NIM)'' (M.C. and P.H.). -- One of us (P.H.) also wishes to acknowledge those many stimulating and inspiring scientific discussions
with Peter G. Wolynes, who is still young enough to appreciate and to contribute great science.

\appendix
\section{Appendix. $f$ dimensional log-oscillators}
Consider a $f$ dimensional log-oscillator:
\begin{align}
H_{\text{log}}(\mathbf X,\mathbf P) = \frac{\mathbf P^2}{2M}+fk_BT\ln\frac{|\mathbf X|}{b}\, ,
\label{eq:Hlogf}
\end{align}
Where $\mathbf X = (X_1, X_2, ... X_f)$, $\mathbf P = (P_1, P_2, ... P_f)$.
For the phase volume $\Phi_\text{log}(E)=\int_{H\leq E} d\mathbf{X}d\mathbf{P}$ one obtains:
\begin{align}
\Phi_\text{log}(E) = \frac{ \left( 8 \pi^2 b^2 Mk_BT/f \right)^{f/2} }{\Gamma(f+1)} e^{E/k_BT}
\end{align}
where $\Gamma$ denotes the Gamma function.
Therefore, the density of states is exponential in $E/k_BT$; reading
\begin{align}
\Omega_\text{log}(E) = \frac{\partial \Phi_\text{log}(E)}{\partial E}=\frac{ \left( 8 \pi^2 b^2 Mk_BT/f \right)^{f/2} }{k_BT\, \Gamma(f+1)} e^{E/k_BT} \;.
\end{align}
Consequently, the methods presented above can also be implemented
with an $f$ dimensional oscillators replacing the 1 dimensional oscillator:
In this case Method I becomes:
\begin{displaymath}
\left\{ \begin{array}{lll}
\dot{\mathbf x} = \mathbf {p}/m, \\
\dot{\mathbf p} = -\partial_{\mathbf x} U(\mathbf x) -\partial_{\mathbf x} h(\mathbf x,X) \\
\dot {\mathbf X} = {\mathbf P}/M\\
\dot {\mathbf P} = -[k_BT/\mathbf X^2]{\mathbf X} -\partial_{\mathbf X}  h(\mathbf x,X)
\end{array}\right.
\end{displaymath}
and Method II becomes
\begin{displaymath}
\left\{ \begin{array}{l}
\dot{\mathbf x} = \mathbf {p}/m+ [k_BT/(\mathbf g-\mathbf X)^{2}] \sum_k (g_k-X_k) \partial_{\mathbf p}\, g_k \\
\dot{\mathbf p} = -\partial_{\mathbf x} U - [k_BT/(\mathbf g-\mathbf X)^{2}] \sum_k (g_k-X_k) \partial_{\mathbf x}\, g_k \\
\dot {\mathbf X} = \mathbf{P}/M\\
\dot {\mathbf P} = [k_BT/ (\mathbf g-\mathbf X)^{2} ](\mathbf g-\mathbf X)
\end{array}\right.
\end{displaymath}
where $\mathbf g$, a short notation for $\mathbf g(\mathbf x, \mathbf p)= (g_1(\mathbf x, \mathbf p), \dots , g_f(\mathbf x, \mathbf p))$, is
an $f$-dimensional field.


\end{document}